\documentclass[twocolumn]{aastex631}

\usepackage{ulem}
\usepackage{hyperref} 
\usepackage{amsmath}
\def\kms{\,\mathrm{km\,s}^{-1}}
\shorttitle{GSE chevron-like substructures}
\shortauthors{Wenbo et al.}
\graphicspath{{./}{figures/}}
\UseRawInputEncoding
\begin{document}

\title{Detection of multiple phase space overdensities of GSE stars by orbit integration}

\correspondingauthor{Gang Zhao}
\email{gzhao@nao.cas.cn}
\author{Wenbo Wu}
\affil{CAS Key Laboratory of Optical Astronomy, National Astronomical Observatories, Chinese Academy of Sciences, Beijing 100101, People's Republic of China; gzhao@nao.cas.cn}
\affil{School of Astronomy and Space Science, University of Chinese Academy of Sciences, Beijing 100049, People's Republic of China}
\affil{Instituto de Astrofísica de Canarias, C/ Vía Láctea s/n, E-38205 La Laguna, Tenerife, Spain}
\author{Gang Zhao}
\affil{CAS Key Laboratory of Optical Astronomy, National Astronomical Observatories, Chinese Academy of Sciences, Beijing 100101, People's Republic of China; gzhao@nao.cas.cn}
\affil{School of Astronomy and Space Science, University of Chinese Academy of Sciences, Beijing 100049, People's Republic of China}
\author{Jiang Chang}
\affil{Purple Mountain Observatory, CAS, No. 10 Yuanhua Road, Qixia District, Nanjing 210034, Peopleʼs Republic of China}
\author{Xiang-Xiang Xue} \affil{CAS Key Laboratory of Optical Astronomy, National Astronomical
	Observatories, Chinese Academy of Sciences, Beijing 100101, People's Republic of China; gzhao@nao.cas.cn} 
\affil{School of Astronomy and Space Science, University of Chinese Academy of Sciences, Beijing 100049, People's Republic of China}
\author{Yuqin Chen} 
\affil{CAS Key Laboratory of Optical Astronomy, National Astronomical
	Observatories, Chinese Academy of Sciences, Beijing 100101, People's Republic of China; gzhao@nao.cas.cn} \affil{School of Astronomy and Space Science, University of Chinese Academy of Sciences, Beijing 100049, People's Republic of China}
\author{Chengdong Li} 
\affil{Université de Strasbourg, CNRS UMR 7550, Observatoire astronomique de Strasbourg, 11 rue de l’Université, 67000 Strasbourg, France}
\author{Xianhao Ye}
\affil{CAS Key Laboratory of Optical Astronomy, National Astronomical Observatories, Chinese Academy of Sciences, Beijing 100101, People's Republic of China; gzhao@nao.cas.cn}
\affil{School of Astronomy and Space Science, University of Chinese Academy of Sciences, Beijing 100049, People's Republic of China}
\affil{Instituto de Astrofísica de Canarias, C/ Vía Láctea s/n, E-38205 La Laguna, Tenerife, Spain}
\author{Chengqun Yang}
\affiliation{Shanghai Astronomical Observatory, Chinese Academy of Sciences, 80 Nandan Road, Shanghai 200030, People’s Republic of China}



\begin{abstract}
In N-body simulations, nearly radial mergers can form shell-like overdensities in the sky position and phase space ($r-v_r$) due to the combination of dynamical friction and tidal stripping. The merger event of Gaia-Sausage-Enceladus has provided a unique opportunity to study the shells in the phase space. To search for them, we integrate the orbits of 5949 GSE-related halo K giants from the LAMOST survey and record their positions at all time intervals in $r-v_r$ diagram. After the subtraction of a smoothed background, we find six significant and complete thin chevron-like overdensities. The apocenters $r_\mathrm{apo}$ of stars in the six chevrons are around 6.75, 12.75, 18.75, 25.25, 27.25, and 30.25 kpc. These chevrons reveal the multiple pile-ups of GSE stars at different apocenters. The application of a different Milky Way mass $M_\mathrm{vir}$ will change the opening angles of these chevrons, while leave their apocenters almost unchanged. By comparing with a recent study of the phase space overdensities of local halo stars from Gaia RVS survey, our results are more inclined to a medium $M_\mathrm{vir}$ of $10^{12}\,M_\odot$. The application of a non-axisymmetric Galactic potential with a steadily rotating bar has a blurring effect on the appearance of these chevron-like overdensities, especially for the chevrons with $r_\mathrm{apo} > 20$ kpc.

\end{abstract}

\section{Introduction} \label{sec:intro}
The arrival of the \textit{Gaia} \citep{2018A&A...616A...1G,2022arXiv220800211G} data reveals that the Galactic stellar halo is dominated by a relatively metal-rich and highly eccentric (eccentricity ec $>$ 0.7) stellar component of a major merger event known as Gaia-Sausage-Enceladus occurred $11-8$ Gyr ago \citep{2018MNRAS.478..611B,2018Natur.563...85H}. GSE is thought to be a massive dwarf galaxy with a total stellar mass $M_*$ on the order of $10^9 - 10^{10}\,M_\odot$ \citep{2019MNRAS.487L..47V,2019MNRAS.490.3426D,2019MNRAS.484.4471F,2019MNRAS.482.3426M}. The stellar debris of the GSE is characterized by the large eccentricity and low angular momentum $L_z$, and the metallicity distribution [Fe/H] of GSE stars peaks around $-1.4 - -1.1$ dex \citep{2019ApJ...881L..10S,2019NatAs...3..932G,2020MNRAS.493.5195D,2020MNRAS.497..109F,2020ApJ...901...48N,2021ApJ...919...66B,2021SCPMA..6439562Z,2022MNRAS.517.2787L,2022ApJ...935..109L,2022ApJ...938...21M}.

In N-body simulations, tidal debris of nearly radial merger events can create patterns similar to shells or umbrellas both in the sky position \citep{2015MNRAS.450..575A,2019MNRAS.487..318K,2022arXiv220808443V} and phase space ($r - v_r$) \citep{1996A&A...310..757S,2013MNRAS.435..378S,2022MNRAS.510..230D}. Such tidal features are found in M31 \citep{2007MNRAS.380...15F} and some elliptical galaxies \citep{2015MNRAS.454.2472H,2022A&A...660A..28B} from their surface brightness. The massive and radial merger event of GSE provides a unique way to search for possible stellar shells in phase space. A recent study using local halo stars from Gaia DR3 RVS survey revealed that stars likely belonging to the GSE form several thin chevron-like overdensities in $r-v_r$ diagram \citep{2023MNRAS.518.6200B}. Further simulations showed that stars in the chevron-like shells conserve similar orbits of unique average energy, and the overdensities in $r-v_r$ space correspond to the bumps of stars in energy distribution \citep{2023MNRAS.518.6200B,2023MNRAS.519..530D}. However, their star sample is constrained to the local stellar halo, and some of the overdensities do not show an overall morphology.

In this letter, we integrate the orbits of our selected GSE stars to find possible chevron-like shells constructed by stars on similar orbits. In previous studies, the orbit integration method is useful to obtain the complete spatial and kinematic distributions of the Galactic stellar halo when the star sample is limited \citep{2022ApJ...927..145S,2022AJ....164..241Y}. In Section~\ref{sec:data}, we describe the selection of the GSE-related stars and the orbit integration method. In Section~\ref{sub:chevron}, we show the detection of multiple thin stellar chevron-like overdensities with the selection effects corrected. We discuss the influence of applying a different virial mass $M_\mathrm{vir}$ and a steadily rotating bar in Section~\ref{sub:potential} and ~\ref{result:bar}. Our results are summarized in Section~\ref{sec:summary}.  


\section{Data Selection and Method} \label{sec:data}
\begin{figure*}
	\centering
	\includegraphics[width = 1.0\textwidth]{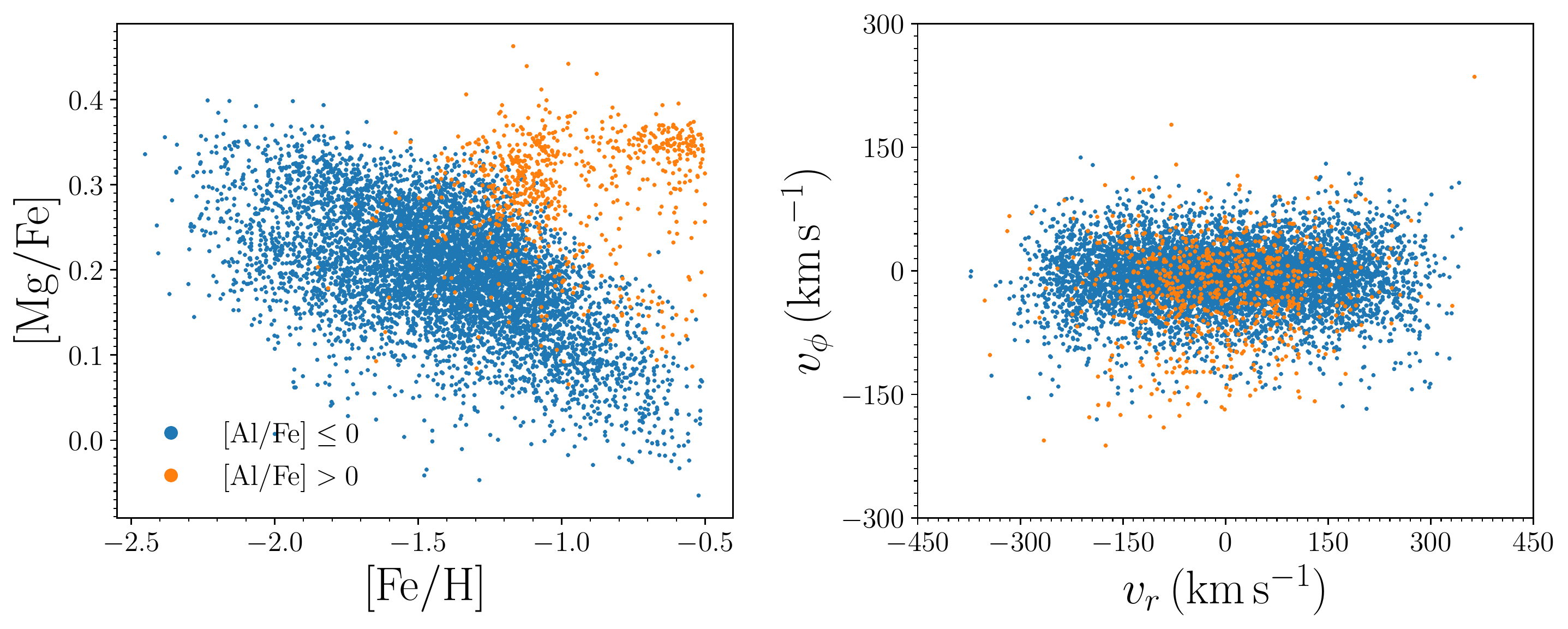}
	\caption{Chemical and kinematic patterns of the K giant sample. Left panel: By setting a boundary of [Al/Fe] = $-0.05$, we can divide these stars into one larger group (blue points, accreted populations of the GSE) and two smaller groups (yellow points, in-situ populations) in [Fe/H] - [Mg/Fe] abundance diagram. Right panel: The accreted group has a typical thin sausage-like pattern in $v_r-v_\phi$ diagram, while the in-situ populations are more dispersed in the distribution of $v_\phi$. }
	\label{fig:ALFEandVelocity}
\end{figure*}

The K giants are selected from the LAMOST DR5 survey \citep{2006ChJAA...6..265Z,2012RAA....12..723Z,2012RAA....12.1197C,2012RAA....12.1243L,2015RAA....15.1089L} and identified by a support vector machine classifier based on the spectral line features \citep{2014ApJ...790..110L}. Distances of these K giants are estimated by a Bayesian method presented in \citet{2014ApJ...784..170X}. By making use of a Gaussian Mixture model, we select around 8000 GSE stars from this K giant sample through the 3D velocities and metallicity [Fe/H] in our previous studies \citep{2022ApJ...924...23W,2022AJ....164...41W}. However, this sample may suffer from the contamination of the in-situ halo \citep{2022AJ....164..249H}. To ensure its purity, we remove the in-situ halo component by requiring [Al/Fe] $< 0$ and signal-to-noise ratio SNR $>$ 20 as done in \cite{2022MNRAS.514..689B}. In Figure~\ref{fig:ALFEandVelocity}, we can see that this criterion divides our star sample into one larger accreted group and two smaller possible in-situ groups in [Fe/H]$-$[Mg/Fe] panel. The accreted group has a clear sausage-like pattern in $v_r-v_\phi$ diagram. Values of [Al/Fe], [Mg/Fe] and SNR are provided by a value added catalog of the LAMOST survey using a deep convolutional neural network \citep{2022MNRAS.517.4875L}  

The orbits are integrated by a python package \texttt{galpy} and a Milky Way potential \texttt{MWPotential2014} \citep{2015ApJS..216...29B}. We use a bigger dark matter halo mass of $M_\mathrm{vir} = 1.0\times10^{12}\,M_\odot$ rather than $0.8\times10^{12}\,M_\odot$. In Galactocentric Cartesian coordinates, we use the values of the Solar Galactocentric distance $r_{\mathrm{gc},\odot}$ = 8.122 kpc \citep{2018A&A...615L..15G}, and height $Z_{\odot}$ = 20.8 pc \citep{2019MNRAS.482.1417B}. We adopt a Solar motion of (+12.9, +245.6, +7.78) km $\textup{s}^{-1}$ \citep{2004ApJ...616..872R,2018A&A...615L..15G,2018RNAAS...2..210D}. The uncertainties in distance, line-of-sight radial velocity, and proper motion (including the measurement error and covariance of proper motions) are propagated using Monte Carlo sampling in order to estimate the median and standard error of the orbital period $P$. We remove stars with large uncertainties in the orbit integration by requiring $\Delta P/P < 0.3$. The final data sample contains 5949 GSE-related K giants. For each star, we integrate 1 Gyr and store 1000 particles at all equal time intervals. Following \citet{2017RAA....17...96L}, the selection function $S$ in which the probability of a K giant being included in the observation is given by their Galactic coordinates ($\textit{l}, \textit{b}$), colors ($\textit{c}$), and apparent magnitudes ($\textit{m}$).  
\section{Results} \label{sec:result}
\subsection{Chevron-like overdensities} \label{sub:chevron}
\begin{figure*}
	\centering
	\includegraphics[width = 1.0\textwidth]{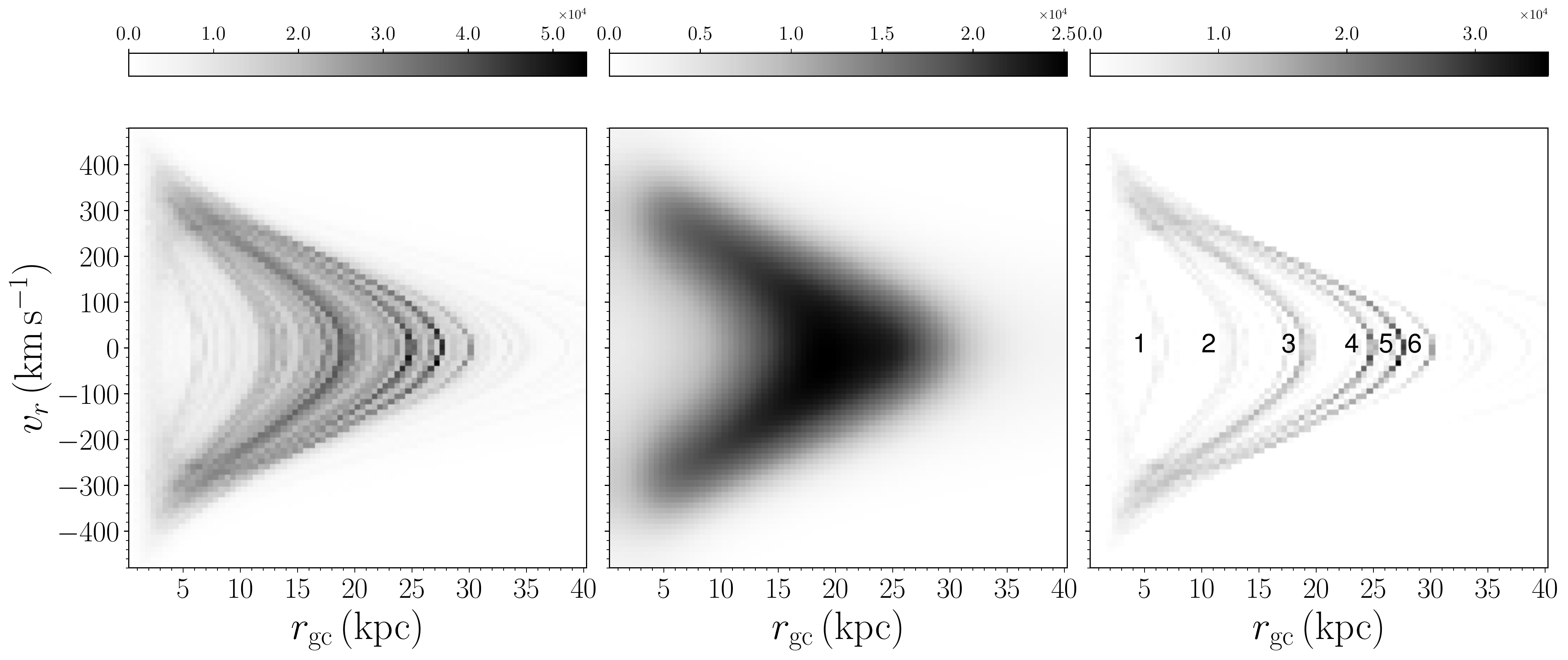}
		\includegraphics[width = 0.5\textwidth]{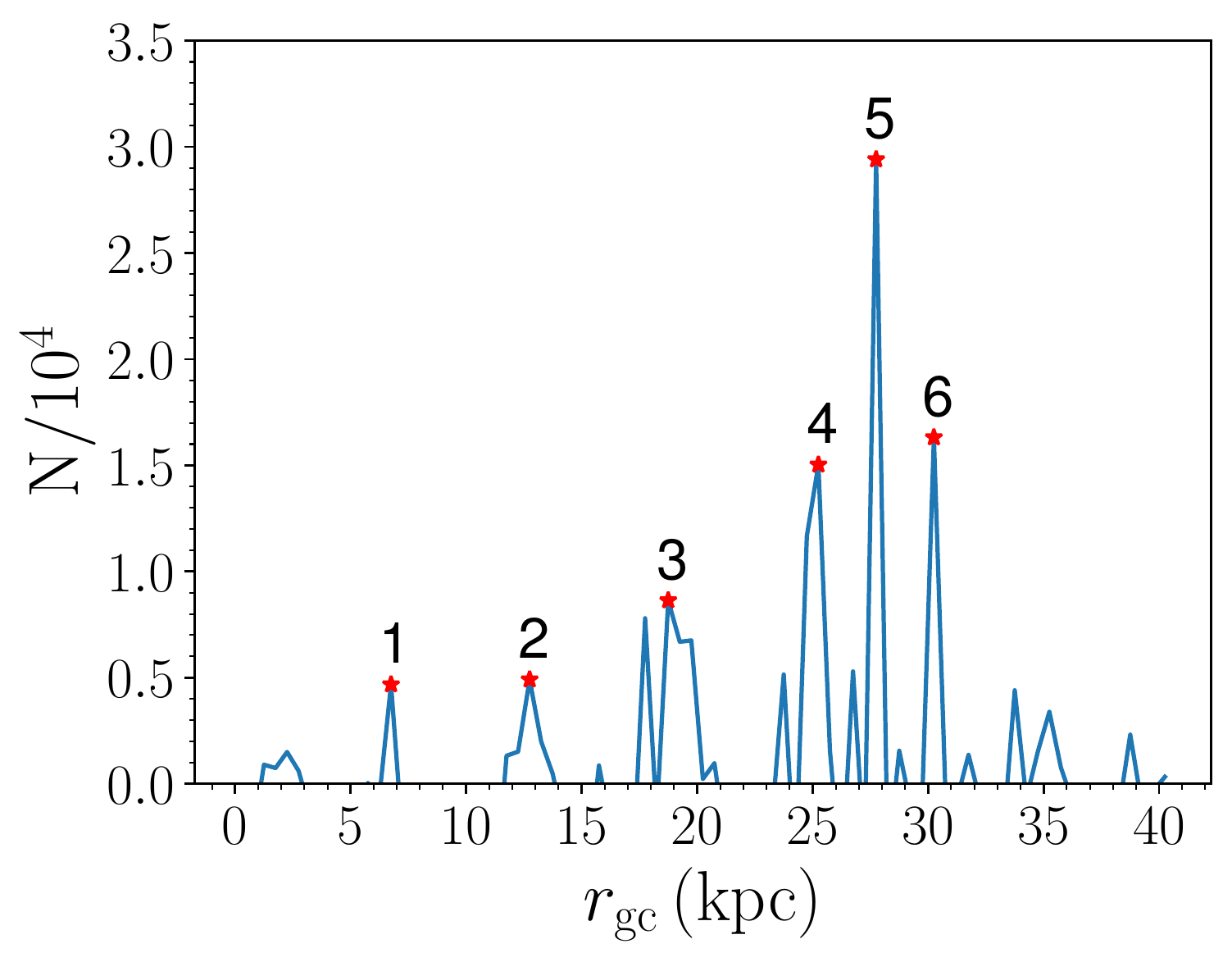}
	\caption{Top left: Stellar particle density $\mathrm{N}$ of the GSE stars in $r_\mathrm{gc}-v_r$ space with selection effects corrected. This diagram contains 81$\times$81 pixels, and each pixel has an equal size of 0.5 kpc$\times$12 $\mathrm{km\,s^{-1}}$. Top middle: Smoothed background obtained by convolving the density image with a Gaussian filter with a FWHM of 5 pixels. Top right: Background subtracted density image with chevron-like overdensities labeled. We set a lower limit of $\mathrm{N} = 0$ to better show the overdensity regions. Bottom: We select pixels of $|v_r| < 6\,\mathrm{km\,s^{-1}}$ from the background subtracted density image and display them as a function of $r_\mathrm{gc}$. We can see several peaks corresponding to the chevron-like overdensities. The apocenters of chevrons 1 to 6 are 6.75, 12.75, 18.75, 25.25, 27.25, and 30.25 kpc.}
	\label{fig:rvshell_1.0}
\end{figure*}

\begin{figure*}
	\centering
	\includegraphics[width=1.0\textwidth]{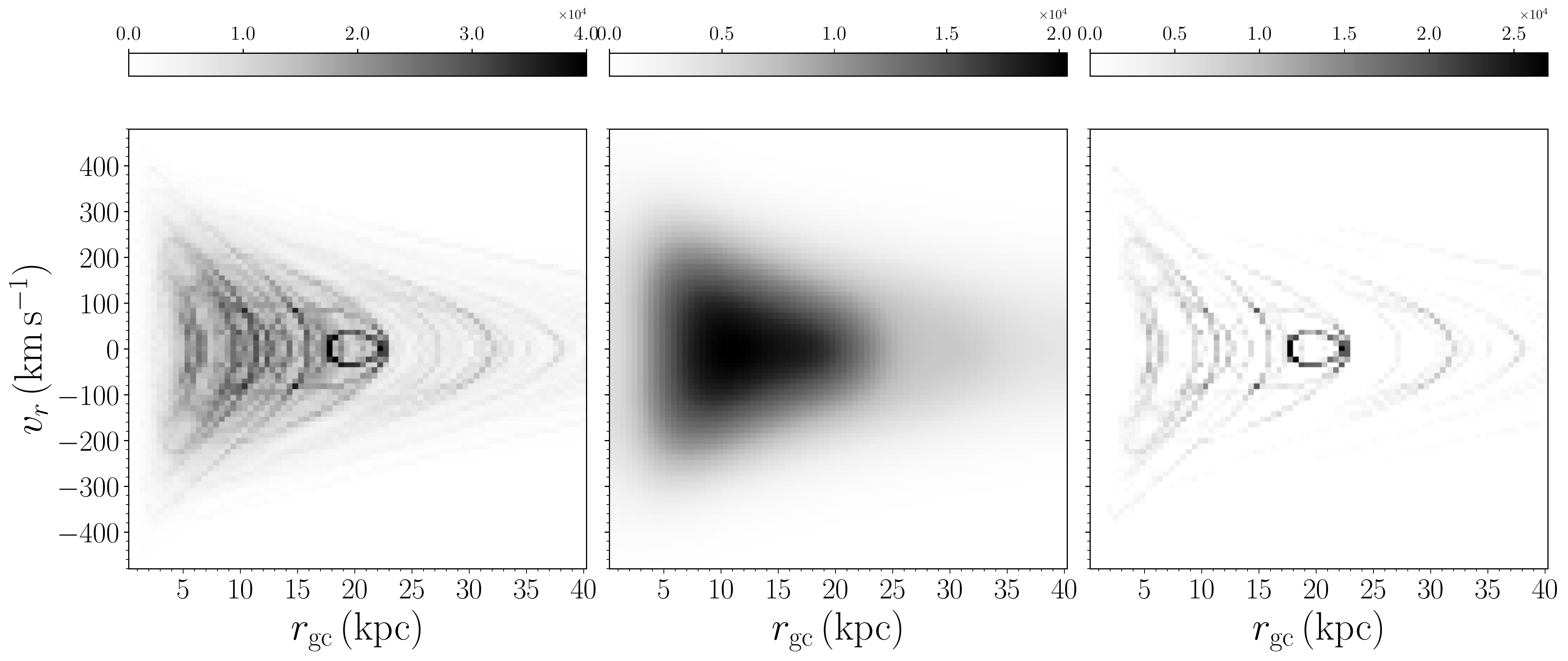}
	\caption{Stellar particle density $\mathrm{N}$ of the GSE-removed stars in $r_\mathrm{gc}-v_r$ space with selection effects corrected. Compared to the GSE stellar particles, they have a much more messy distribution in $r_\mathrm{gc}-v_r$ space. After the subtraction of a smoothed background, the overdensity regions tend to overlap with each other rather than form several separated chevrons as Figure~\ref{fig:rvshell_1.0}. These overdensity regions maybe related to the existence of multiple minor halo substructures in the K giant sample identified by Xue et al. (in preparation).} 
	\label{fig:GSE-removed}
\end{figure*}

In the left panel of Figure~\ref{fig:rvshell_1.0}, we show the density distribution of all particles in $r-v_r$ map. Each particle of a star $i$ is weighted by the selection function $S_i$. We use a two-dimensional Gaussian smoothing filter \texttt{scipy.ndimage.gaussian\_filter} to obtain a smoothed distribution as a background. In the right panel of Figure~\ref{fig:rvshell_1.0}, six thin and complete chevron-like overdensities are clearly visible after the subtraction of a smoothed background. These overdensities are constructed by GSE stars on similar orbits. As a comparison, we also show the density distribution of the GSE-removed stellar particles in Figure~\ref{fig:GSE-removed}, including the GSE-removed stars selected from \citet{2022AJ....164...41W} and stars satisfying [Al/Fe] $>$ 0 in Figure~\ref{fig:rvshell_1.0}. After the subtraction of a smoothed background, the overdensity regions overlap with each other rather than form several clearly separated chevrons like Figure~\ref{fig:rvshell_1.0}.

In a nearly spherical potential, the trace of a star orbit in $r-v_r$ space is determined by two free parameters of apocenter $a$ and eccentricity ec. The eccentricity of GSE stars in different overdensities is similar (ec $>$ 0.8), so the main difference is their apocenters. To obtain the apocenters, we extract values of the background subtracted distribution at $v_r\sim0$ (pixels satisfying $|v_r| < 6\,\mathrm{km\,s^{-1}}$) and display them as a function of $r_\mathrm{gc}$ in the bottom panel of Figure~\ref{fig:rvshell_1.0}. We can see several clear pulse signals caused by the pile-up of particles of the six overdensities. We define the apocenter of these overdensities as the corresponding $r_\mathrm{gc}$ of the peak of the pulse signals. The apocenters of the six overdensities are 6.75, 12.75, 18.75, 25.25, 27.25, and 30.25 kpc, and we label them as chevron 1$-$6. Besides the target selection in the spectroscopic survey, a selection criterion of $|Z| > 2$ kpc is applied to exclude possible disk stars in our K giant sample. This additional selection has a larger impact on stars with a smaller $r_\mathrm{gc}$, which could artificially reduce the strength of chevrons in the inside halo, specifically chevron 1 and 2. Besides these obvious chevrons, we also find two peaks at $r_\mathrm{gc} = 35$ and 39 kpc. Since they are too weak and not complete in the background subtracted distribution image, we are not sure about their reliabilities.

These chevron-like overdensities reveal the multiple pile-ups of GSE stars at different apocenters. \citet{2022AJ....164..249H} find two breaking radii at $r_\mathrm{gc}$ of 12 and 28 kpc when fitting the stellar density distribution of the accreted halo of the H3 survey with a multiply broken power law. The apocenters of chevron 2 and 5 closely coincide with the two breaking radii, and this consistency has been predicted by the N-body simulation of \citet{2021ApJ...923...92N}. A series of studies argue that the major merger of GSE is likely to be composed of multiple smaller mergers with different chemodynamics including the apocenters \citep{2020ApJ...902..119D,2022ApJ...932L..16D,2022arXiv221112576D}. Chevron 1 is possibly related to their identified substructure Cronus with a mean apocenter of $7.7\pm2.2$ kpc, and chevron 2 has a similar apocenter as the substructures of Nereus and Virgo Radial Merger. The difference between these substructures is mainly shown in the element abundance space. However, the low-resolution spectroscopic survey of our K giants sample prevents us from a comprehensive study of the chemical patterns of stars inside these chevrons.

\subsection{Influence of a different potential and constrain of virial mass} \label{sub:potential}
\begin{figure*}
	\centering
	\includegraphics[width = 1.0\textwidth]{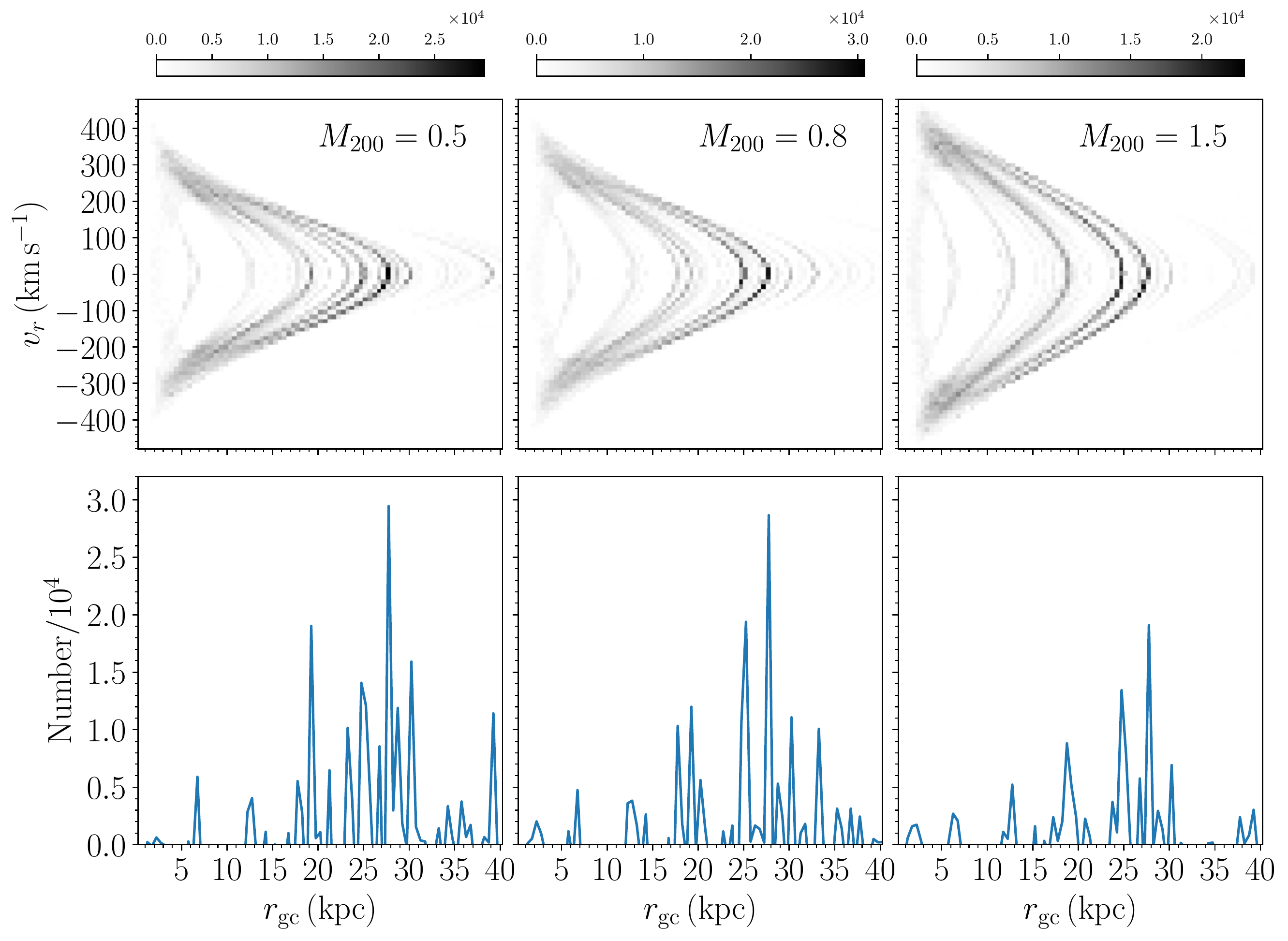}
	\caption{Background subtracted density image (top row) and values of pixels satisfying $|v_r| < 6\,\mathrm{km\,s^{-1}}$ (bottom row) for $M_\mathrm{vir} = 0.5\times10^{12}\,M_\odot$ (left column), $M_\mathrm{vir} = 0.8\times10^{12}\,M_\odot$ (middle column), and $M_\mathrm{vir} = 1.5\times10^{12}\,M_\odot$ (right column). The main changes of chevron 1 to 6 are the strength and opening angles, while their apocenters keep almost constant.}
	\label{fig:otherPotential}
\end{figure*}

We adopt a virial mass of $M_\mathrm{vir} = 10^{12}\,M_\odot$ in the above orbit integration. However, large uncertainties still exist in the current estimation of the Milky Way mass. Using different methods and tracers, values of the estimated $M_\mathrm{vir}$ range from a lower side of $0.5\times10^{12}\,M_\odot$ to an upper side of $1.5\times10^{12}\,M_\odot$. To check its influence, we repeat the above steps and obtain the background subtracted distributions with a different $M_\mathrm{vir}$. In this paper, we adopt four values of $M_\mathrm{vir}$, which are $0.5\times10^{12}\,M_\odot$ (K giants, \citet{2022MNRAS.516..731B}), $0.8\times10^{12}\,M_\odot$ (giants, \citet{2014ApJ...794...59K}), $10^{12}\,M_\odot$ (BHB stars, \citet{2022MNRAS.516..731B}), and $1.5\times10^{12}\,M_\odot$ (globular clusters, \citet{2019ApJ...873..118W}). Combing Figure~\ref{fig:rvshell_1.0} and ~\ref{fig:otherPotential}, we can see that the main changes in chevron 1$-$6 are the morphology and strength, while their apocenter keeps almost constant. For the six chevrons, their opening angles decrease and strength increases with a smaller $M_\mathrm{vir}$. Besides them, there are another two possible chevron-like overdensities with apocenters of 23.25 and 40 kpc in $M_\mathrm{vir} = 0.5\times10^{12}\,M_\odot$. 

The formation of these chevrons is possibly related to the non-uniform energy distribution of the GSE stars. By studying the N-body stellar particles of GSE stars, \citet{2023MNRAS.518.6200B} found prominent bumps related to the leading and trailing arms in the energy distribution. Besides, they also found smaller-scale bumps that become smoothed and fade away with time. Each stripping episode could cause at least two bumps, and the accretion event of GSE is thought to have three main stripping episodes. Therefore, we could expect more than six chevron-like overdensities related to the energy bumps. Besides the chevrons in this study, \citet{2023MNRAS.518.6200B} also find signatures of other five chevron-like overdensities in $r-v_r$ diagram at $r_\mathrm{gc} = 4$ to 12 kpc using Gaia DR3 RVS survey. Through a linear extrapolation method, they estimate that the apocenters of the five overdensities are 11.5, 15.5, 21, 23, and 25 kpc. Here we label them as Be 1$-$5. \citet{2023MNRAS.518.6200B} do not need an orbit integration method due to the large number of halo stars in a small spatial region, thus means their results are independent of the Galactic potential. However, their star sample is constrained to the local stellar halo and may suffer from the contamination of the in-situ halo. Our GSE star sample has a wider distribution of $r_\mathrm{gc}$ and shows a complete morphology of the chevron-like overdensities, but this result is highly potential-dependent. Therefore, a comparison between our study and \citet{2023MNRAS.518.6200B} not only helps in identifying the overdensities but also in constraining the range of $M_\mathrm{vir}$. 

\begin{figure*}
	\centering
	\includegraphics[width = 0.8\textwidth]{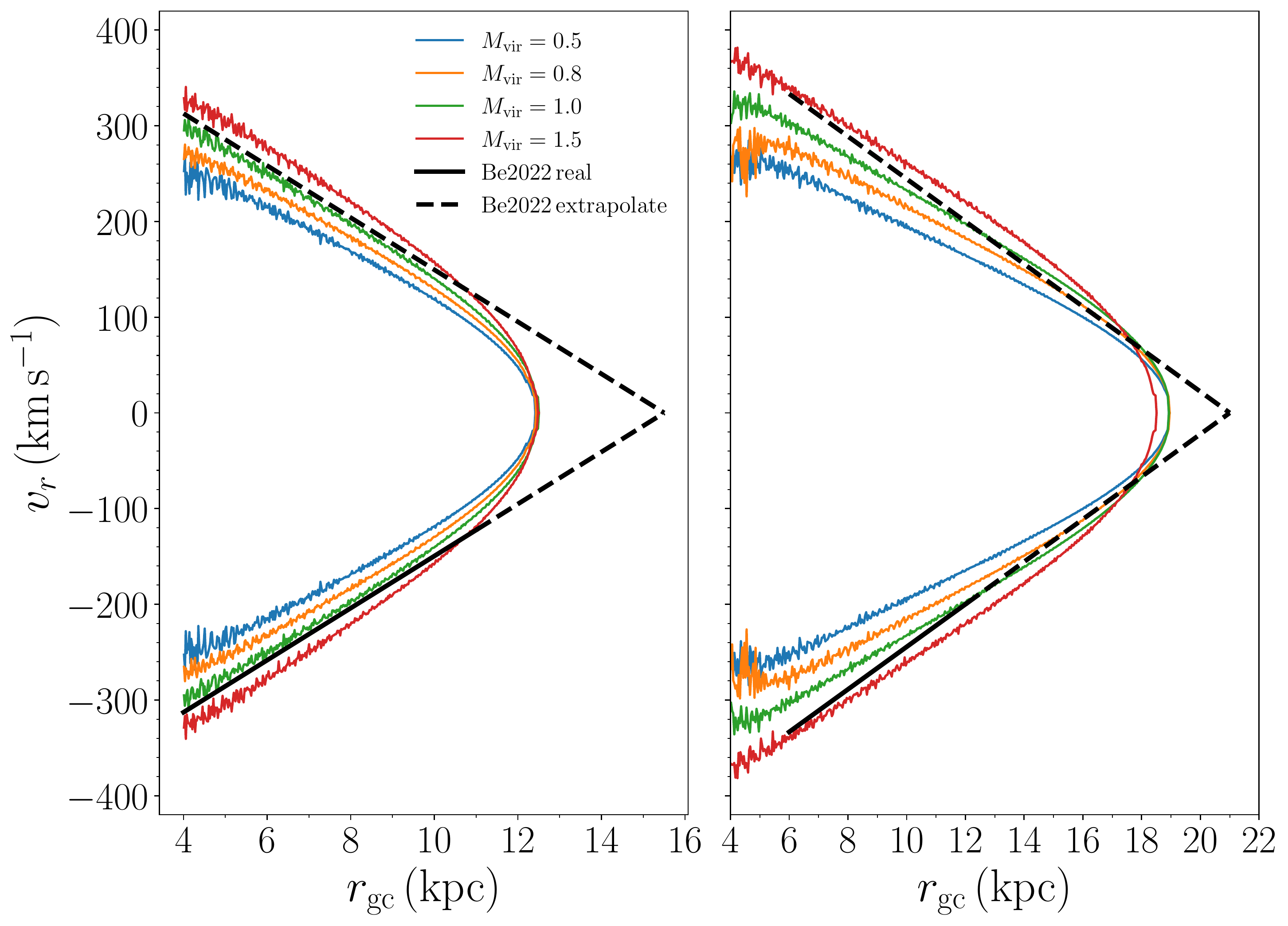}
	\caption{Left panel shows a comparison between Be 2 and chevron 2 of different $M_\mathrm{vir}$. We select stars of the same apocenter as chevron 2. The approximate morphology of chevron 2 is obtained by averaging the positions of these stellar particles in $r-v_r$ space. The black solid line is the approximate location of Be 2 in the real observation, while the dash line is the linear extrapolation result. Chevron 2 of $M_\mathrm{vir} = 10^{12}\,M_\odot$ (green solid line) agrees well with Be 2. Right panel shows a comparison of chevron 3 and Be 3. Be 3 is located between chevron 3 of $M_\mathrm{vir} = 10^{12}\,M_\odot$ (green solid line) and $M_\mathrm{vir} = 1.5\times10^{12}\,M_\odot$ (red solid line).}
	\label{fig:CompareBe2022}
\end{figure*}

The signatures of Be 4 and 5 are too weak, and Be 1 has no corresponding chevron in our results. Therefore, we compare Be 2 and Be 3 with chevron 2 and chevron 3 respectively in Figure~\ref{fig:CompareBe2022}. Be 2 is very consistent with chevron 2 of $M_\mathrm{vir} = 10^{12}\,M_\odot$ at $r_\mathrm{gc} = 4-10$ kpc. At $r_\mathrm{gc} > 12$ kpc, we can see that Be 2 of the linear extrapolation deviates from chevron 2, and apocenter of Be 2 is likely to be overestimated. Be 3 is located between chevron 3 of $M_\mathrm{vir} = 10^{12}\,M_\odot$ and $1.5\times10^{12}\,M_\odot$, and we can hardly to decide which $M_\mathrm{vir}$ is more suitable for Be 3. In general, this comparison is more inclined to a virial mass of $M_\mathrm{vir}=10^{12}\,M_\odot$.

\subsection{influence of a steadily rotating bar}\label{result:bar}
In the last subsection we use an axisymmetric potential during the orbit integration. However, considering the small pericentre ($r_\mathrm{peri} < 5$ kpc) of our K giants and the massive long bulge/bar structure $\sim10^{10}\,M_\odot$ \citep{2015MNRAS.448..713P,2017MNRAS.465.1621P}, the integrated orbits could be heavily perturbed by the bar when passing the Galactic centre or resonance trapped regions related to the bar. Previous studies show that the dynamical influence of a rotating bar is not only constrained to the disc stars \citep{2022ApJ...936L...7C,2022MNRAS.513..768C,2023arXiv230306393L} but also reaches to a much further region \citep{2018AstBu..73..162C}. The bar-like perturbations is possibly related to the creation and growth of several substructures, such as Hyades, Sirius, Hercules, and Ophiuchus streams \citep{1998AJ....115.2384D,2014A&A...563A..60A,2016MNRAS.460..497H,2019MNRAS.484.4540H}. A recent simulation work of \citet{2023MNRAS.521L..24D} showed that a rotating bar could accelerate the phase-mixing of these chevron-like substructures found in \citet{2023MNRAS.518.6200B}.

We add a steadily rotating bar to the modified \texttt{MWPotential2014} with $M_\mathrm{200} = 10^{12}\,M_\odot$ by a python package \texttt{Agama} \citep{2019MNRAS.482.1525V}. To represent the bar, we choose a model following \citet{2022MNRAS.513..768C} as
\begin{equation}
	\Phi_\mathrm{b}(r, \theta, \phi, t) = \Phi_\mathrm{br}(r)\sin^2\theta\cos m(\phi - \Omega_\mathrm{b}t),
\end{equation}  
where $(r, \theta, \phi)$ are the spherical coordinates and $m = 2$ is set to only consider the quadrupole term. The radial dependence of the bar potential $\Phi_\mathrm{br}$ is
\begin{equation}
	\Phi_\mathrm{br}(r) = -\frac{A v^2_\mathrm{c}}{2}(\frac{r}{r_\mathrm{CR}})^2(\frac{b+1}{b+r/r_\mathrm{CR}})^5,
\end{equation} 
where $A$ is the strength of the bar, $v_\mathrm{c}$ is the circular velocity in the solar vicinity, and $b$ is the ratio between the bar scale length and the value of co-rotation radius $r_\mathrm{CR}$. Following \citet{2022MNRAS.513..768C}, we set $v_\mathrm{c} = 235\,\kms$, $A = 0.02$, $b = 0.28$, and $r_\mathrm{CR} = 6.7~\rm{kpc}$. The initial phase angles of the bar $\phi_\mathrm{b}$ is $28^\circ$ \citep{Wegg2015}. Different pattern speeds of $\Omega_\mathrm{b} = {-35, -40, -45}~\kms~\rm{kpc}^{-1}$ are chosen in the orbit integration.

\begin{figure*}
	\centering
	\includegraphics[width = 1.0\textwidth]{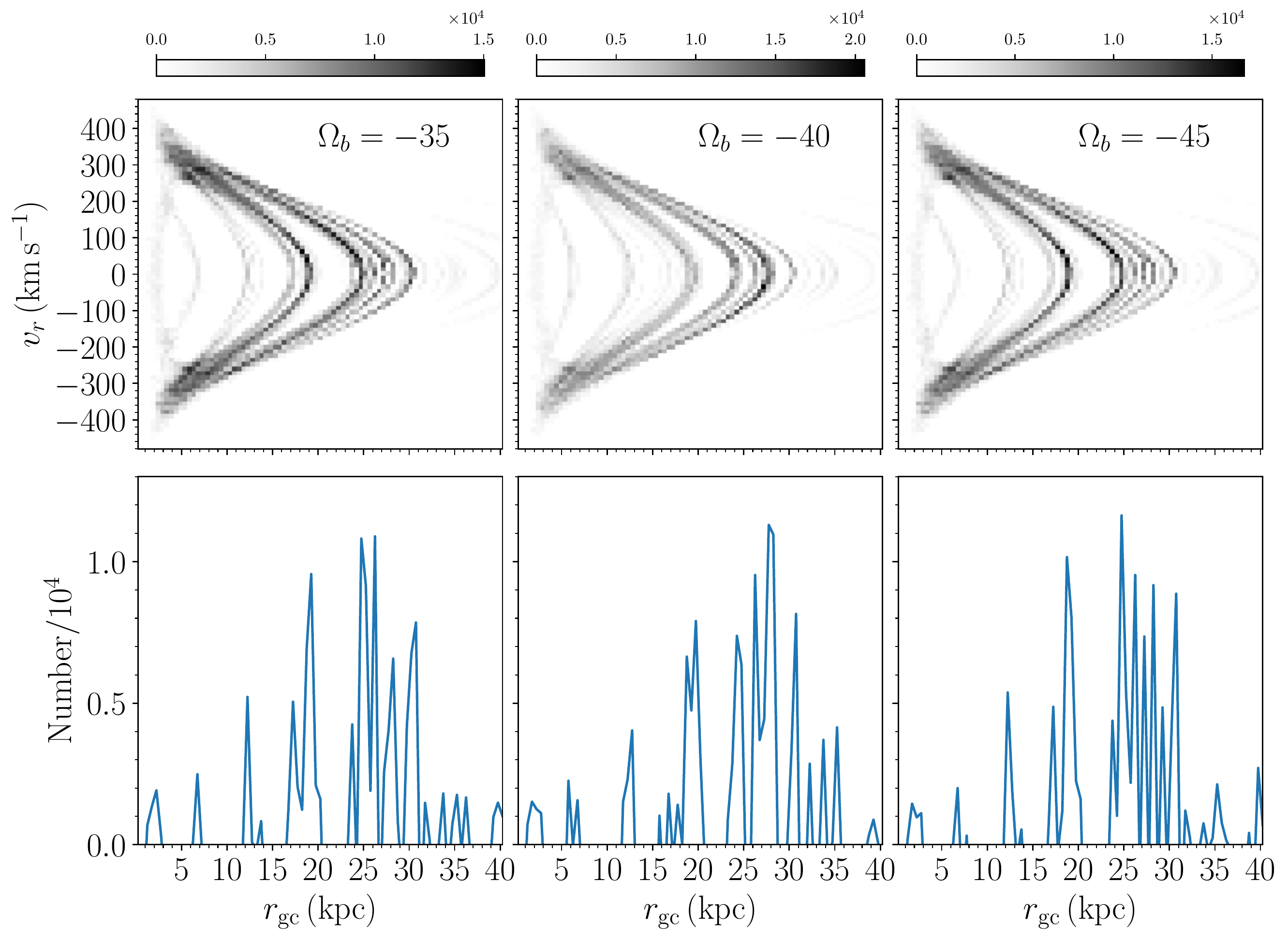}
	\caption{Background subtracted density image (top row) and values of pixels satisfying $|v_r| < 6\,\mathrm{km\,s^{-1}}$ (bottom row) for a bar added potential with $\Omega_\mathrm{b} = -35$ (left column), $\Omega_\mathrm{b} = -40$ (middle column), and $\Omega_\mathrm{b} = -45$ (right column). The steadily rotating Galactic bar has a blurring effect on these overdensities, especially for the chevrons at $r_\mathrm{gc}\sim25-30$ kpc.}
	\label{fig:patternspeed}
\end{figure*}

Figure~\ref{fig:patternspeed} shows the background subtracted images and the distribution of $N$ as a function of $r_\mathrm{gc}$ with different $\Omega_\mathrm{b}$. We find that the introduction of a steadily rotating bar could smooth the overdensity regions, which leads to a decline of the strength of these chevrons. The peak values $N$ of the chevron 4, 5, 6 are 1.5, 2.9, 1.6$\times10^4$ in Figure~\ref{fig:rvshell_1.0}, while these values fall below $1.2\times10^4$ in a bar-added non-axisymmetric potential. 
Our results are consistent with \citet{2023MNRAS.521L..24D} that a rotating bar would result in a blurry chevron at $r_\mathrm{gc} > 20$ kpc.

\begin{figure*}
	\centering
	\includegraphics[width = 0.7\textwidth]{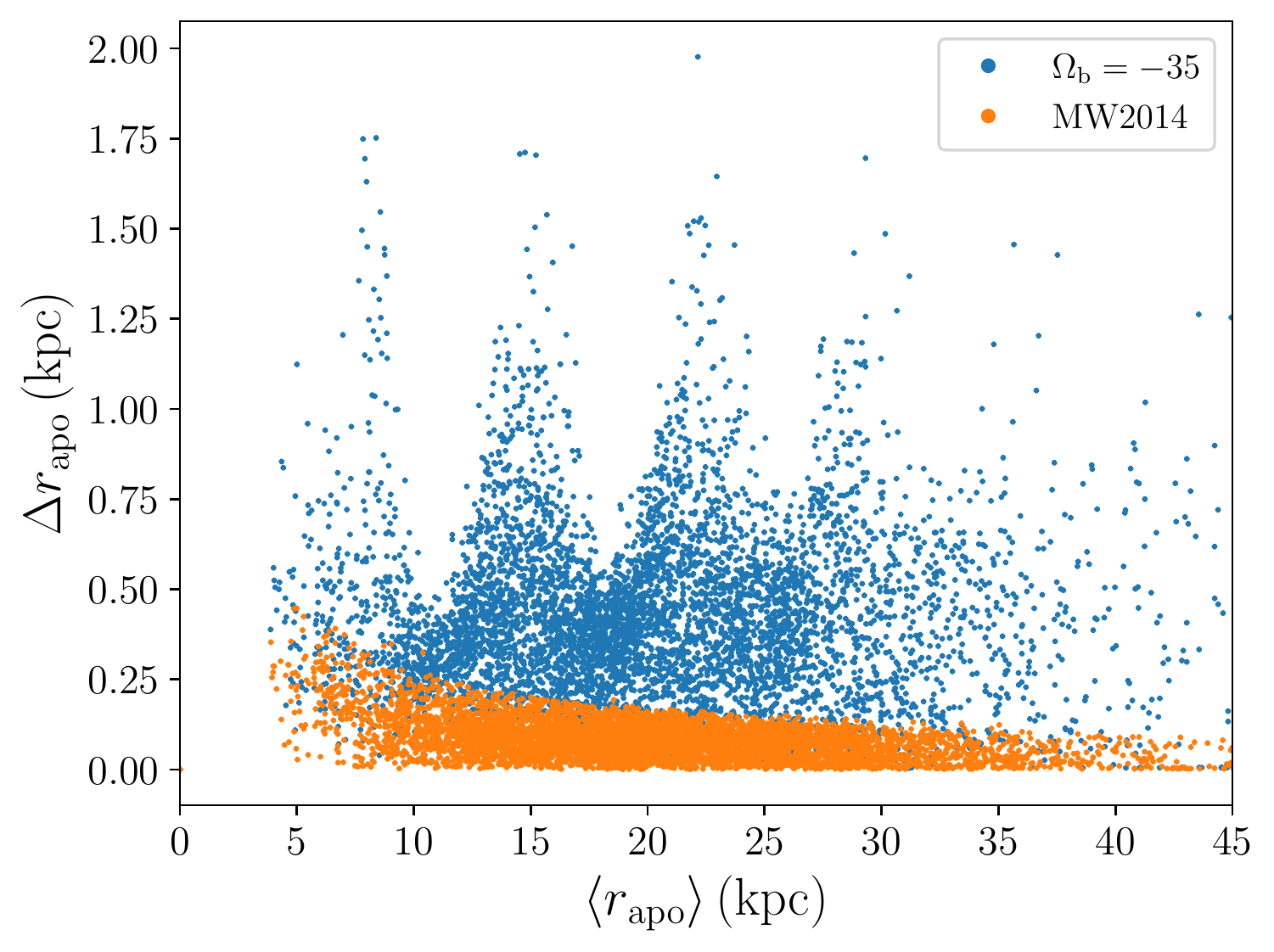}
	\caption{Change of apocenter $\Delta r_\mathrm{apo}$ versus mean value $\langle r_\mathrm{apo}\rangle$ for the GSE stars in the \texttt{MW2014} (orange points) and bar added potential (blue points). The large increase of $\Delta r_\mathrm{apo}$ shows the strong chaotization effect of the Galactic bar on the stellar orbit during the integration.}
	\label{fig:apocenter}
\end{figure*}

The smoothing effect revealed in Figure~\ref{fig:patternspeed} is mainly caused by the disturbance of the stellar orbit under the perturbation of the Galactic bar. In general, these K giants would finish three to ten orbit periods in the 1 Gyr integration. We obtain the apocenters $r_\mathrm{apo}$ of a star $i$ at different periods, and show the difference value $\Delta r_\mathrm{apo}$ ($\mathrm{max}(r_\mathrm{apo}) - \mathrm{min}(r_\mathrm{apo})$) as a function of the mean value $\langle r_\mathrm{apo} \rangle$ in Figure~\ref{fig:apocenter}. The change of $r_\mathrm{apo}$ (a mean value of 0.08 kpc) is almost negligible in a symmetric static potential, while the adding of a rotating bar would cause a large bump of $\Delta r_\mathrm{apo}$ (a mean value of 0.44 kpc). The disturbance leads to a more diffuse distribution of the integrated stellar particles in $r-v_r$ space, which decreases the strength of these chevrons in Figure~\ref{fig:patternspeed}.

Motivated by \citet{2023MNRAS.521L..24D}, we divide the K giant sample into two subsamples of large ($r_\mathrm{peri} > R_\mathrm{b}$, 1648 stars) and small ($r_\mathrm{peri} < R_\mathrm{b}$, 4301 stars) pericentre, where $R_\mathrm{b} = 1.87$ kpc is the bar scale length. Stars with a smaller pericentre are more likely to be strongly influenced by the bar. These chevron-like overdensities are clearly seen in the subsample with small pericentre, while they have a smaller opening angle and a more diffusely mixed appearance in the other subsample. The most significant difference is the chevron 2 ($r_\mathrm{apo}\sim12.75$ kpc), which is obvious in the subsample with small pericentre but almost invisible in the subsample with large pericentre. \citet{2023MNRAS.521L..24D} found that all the chevrons discovered in the Gaia RVS sample almost disappear when they only considered stars with large pericentre ($r_\mathrm{peri} > 2$ kpc). They thought it is somewhat in conflict with the smoothing effect of the bar in simulations, and suspected that a resonance effect may account for the formation of these chevrons. Using a test particle simulation and starting with a smooth stellar halo, \citet{2023arXiv230300008D} found that a wedged-shape overdensity with a tip at 10.5 kpc forms in $r-v_r$ space after the growth of the bar. Therefore, we note that a bar-driven mechanism may work in the formation of these chevron-like overdensities, especially for the chevron 2.

\begin{figure*}
	\centering
	\includegraphics[width = 0.7\textwidth]{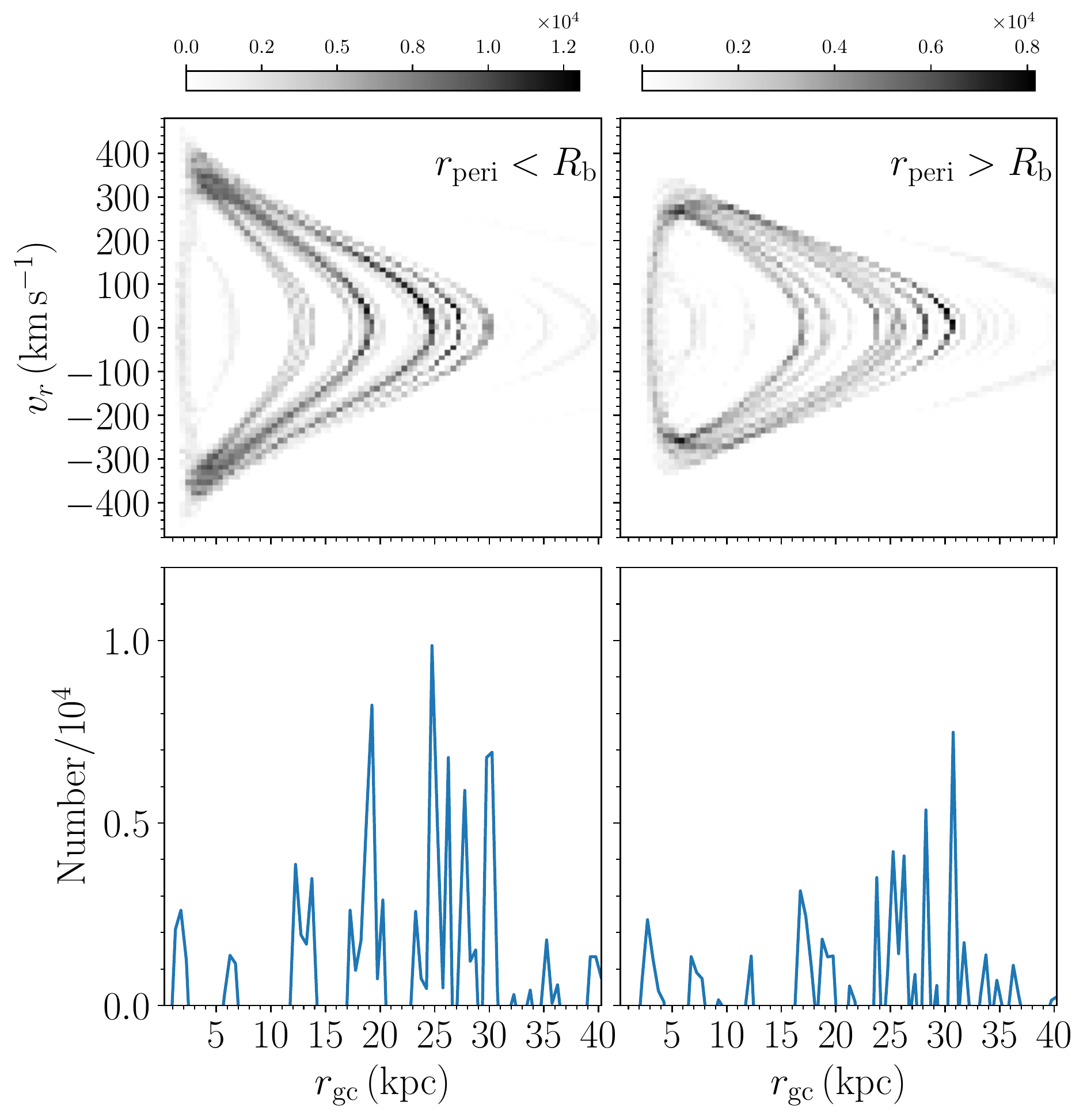}
	\caption{Background subtracted density image (top row) and values of pixels satisfying $|v_r| < 6\,\mathrm{km\,s^{-1}}$ (bottom row) for the subsamples with small pericentre (left column) and large pericentre (right column) in the bar added potential $\Omega_\mathrm{b} = -35$. We can see a very clear image of these chevron-like overdensities in the left hand column. Most of the chevrons can still be found in the right column, but with a smaller opening angle and a more diffusely mixed appearance. Chevron 2 almost disappears when we only consider stars satisfying $r_\mathrm{peri} > R_\mathrm{b}$.}
	\label{fig:rperi}
\end{figure*}

Only the steadily rotating bar model is adopted as the perturbation to the background potential in this work. This scenario reveals two possible flaws in the interpretations within this subsection. On one hand, as it is reported in the literature \citep{Chiba2021}, the pattern speed of the bar decreases in the Galaxy, which leads to an increasing of the scale length \citep{Athanassoula1992,Chiba2021}. 
Thus, the number of orbits strongly impacted by the bar should also increase, which alters the results shown in Figure~\ref{fig:rperi}. Meanwhile, the co-rotation radius of the bar also changes with the change of the pattern speed. Naturally, it's worthwhile to investigate the influence to the shape of the chevron-like structure by the time-varying bar model. On the other hand, the self-gravity is neglected throughout our particle simulation. However, since the stars interested in this work mainly belong to halo population with long orbital periods, the self-gravity could play an important role in shaping the characteristics of the chevron-like structures. This is consistent with the comparison between our simulation and the result from \citet{2023MNRAS.521L..24D}, which shows more blurring chevron structures compared with our simulations. As a conclusion, a more realistic bar model including the decrease of pattern speed together with the self-gravity of the system should be considered in the future to investigate the chevron like features in the Galaxy.

\section{Summary} \label{sec:summary} 
In this study, we explore the stellar distribution of GSE-related K giants of LAMOST survey in $r-v_r$ space by orbit integration. We summarize our results as follows:

1. We find six thin and complete chevron-like overdensities in $r-v_r$ space. The apocenters of these overdensities are 6.75, 12.75, 18.75, 25.25, 27.25 and 30.25 kpc. These chevrons are possible related to the pile-ups of GSE stars at different apocenters. 

2. The application of a different Milky Way mass $M_\mathrm{vir}$ will change the morphologies and strength of these chevrons, while the apocenters keep almost constant.

3. By comparing with a recent potential independent study, we think that our results are more inclined to a $M_\mathrm{vir}$ of $10^{12}\,M_\odot$.

4. The adding of a steadily rotating bar would significantly chaotize the stellar orbit, which leads to a more diffusely mixed appearance of these chevrons especially at $r_\mathrm{gc} > 20$ kpc.

\begin{acknowledgements}
This study is supported by the National Natural Science Foundation of China under grant Nos. 11988101, 11890694, 11873052, 12273027, the National Key R$\&$D Program of China No. 2019YFA0405500 and the International Partnership Program of CAS Grant no. 178GJHZ2022040GC. W.B. W. and X.H. Y acknowledges the support from the China Scholarship Council. We thank the anonymous reviewer for the valuable suggestions. We thank Prof Rafael Rebolo, Prof Carlos Allende, Dr Jonay Gonzalez, Dr Wenyu Xin, and Dr Zhicun Liu for their help. Guoshoujing Telescope (the Large Sky Area Multi-Object Fiber Spectroscopic Telescope LAMOST) is a National Major Scientific Project built by the Chinese Academy of Sciences. Funding for the project has been provided by the National Development and Reform Commission. LAMOST is operated and managed by the National Astronomical Observatories, Chinese Academy of Sciences. This work presents results from the European Space Agency (ESA) space mission Gaia. Gaia data are being processed by the Gaia Data Processing and Analysis Consortium (DPAC). Funding for the DPAC is provided by national institutions, in particular the institutions participating in the Gaia MultiLateral Agreement (MLA).
\end{acknowledgements}

\bibliography{sample631}{}
\bibliographystyle{aasjournal}

\end{document}